\documentstyle[12pt]{article} \newtheorem{theorem}{Theorem}[section]
 
\newtheorem{lemma}[theorem]{Lemma} 
 
\newcommand{\proof}{\noindent {\bf Proof. }} \newcommand{\qed}{\hfill $\Box$ \vskip 2ex}
\renewcommand{\baselinestretch}{1.5} \newcounter{acount}

\newfont{\BB}{msbm10} \def\R{\mbox{\BB R}} \def\C{\mbox{\BB C}} \def\D{\mbox{\BB D}}
\renewcommand{\baselinestretch}{1.5}
\begin{document} 
\title{
Derivation of the wave function collapse in the context of Nelson's stochastic
mechanics}
\author{Michele Pavon\thanks{
Dipartimento di  Elettronica ed
informatica, Universit\`a di  Padova, via Gradenigo 6/A,
and LADSEB, CNR,
35131 Padova, Italy, ph.: 39 049 827 7604, FAX: 39 049 827
7699, E-mail: {\tt pavon@dei.unipd.it.}}} 
\maketitle
\begin{abstract}The von Neumann collapse of the quantum mechanical wavefunction after a
position measurement is derived by a purely probabilistic mechanism in the context of
Nelson's stochastic mechanics. 
\end{abstract}

{\bf Running Title:} Stochastic mechanics and measurement


\newpage

\section {Introduction}
Nelson's stochastic mechanics \cite{Fe,N1,G,N2,BCZ} is a quantization procedure for classical
dynamical systems based on stochastic processes of the diffusion type. This theory leads to
predictions that agree with those of standard quantum mechanics and are confirmed by
experiment. The fundamental assumption is that interaction with a background field causes the
system to undergo a diffusion process with diffusion coefficient $\frac{\hbar}{m}$. A
fascinating hypothesis concerning the origin of the underlying Brownian motion has been recently
advanced by Francesco Calogero in \cite{Cal}. Namely, that this ``tremor" may be caused by
the interaction of every particle with the gravitational force due to all other particles of
the Universe. Following this idea, he obtains a formula for Planck's action constant $h$. The
latter yields the correct order of magnitude for $h$ when current cosmological data are
employed.

It is hardly surprising that the most controversial issue in stochastic mechanics is
the measurement problem. Indeed, in
\cite{G94}, Francesco Guerra writes: ``Therefore, we see that the basic problem in the
interpretation of stochastic mechanics is related to the basic problem in the interpretation
of quantum mechanics: To evaluate the effects of the measurement and explain the mechanism of
the wave packet reduction".

The purpose of this paper is to show that, in the frame of Nelson's stochastic mechanics,
{\it the wave function reduction after a position measurement may be obtained through a purely
probabilistic mechanism}, namely a stochastic variational principle. The
latter has the appealing interpretation of changing the pair of forward and backward drifts of the
reference process as little as possible given the result of the measurement. This
variational principle is  quite similar to the one that yields the new stochastic model after
measurement for nonequilibrium thermodynamical systems, see Section 5, the only difference being
that, in view of the time-reversibility of stochastic mechanics, a time-symmetric kinematics has
to be employed. As we have shown elsewhere \cite{P1,P2,P3}, this kinematics  also permits to develop in
a natural way a Lagrangian and a Hamiltonian formalism in stochastic mechanics. In particular, it
permits to define a {\em momentum process} having the same first and second moment of the
corresponding quantum momentum operator. It is then possible to derive a stochastic counterpart of
Hamilton's canonical equations, and to obtain a simple probabilistic interpretation of the uncertainty
principle \cite{P2} along the lines of \cite{Fe,DLP,DDD,Go}.

\section{Kinematics of finite-energy diffusions} In this section, we review some essential
concepts and results of the kinematics of diffusion processes. We refer the reader to \cite{N1}-
\cite{N3}, \cite {G,F,F2,KS} for a thorough account. Let $(\Omega,{\cal E},{\bf P})$ be a
probability space, and let $I_n$ denote the $n\times n$ identity matrix. 
A stochastic process $\{\xi(t);t_0\le t\le t_1\}$ mapping $[t_0,t_1]$ into
$L^2_n(\Omega,{\cal E},{\bf P})$ is called a {\it finite-energy diffusion} with constant
diffusion coefficient $I_n\sigma^2$ if the increments admit the representation

\begin{equation}\label{K1} \xi(t)-\xi(s)=\int_s^t\beta(\tau)d\tau+\sigma [w_+(t)-w_+(s)],\quad
t_0\le s<t\le t_1, \end{equation} where the {\it forward drift} $\beta(t)$ is at each time $t$ a
measurable function of the past $\{\xi(\tau);0\le \tau\le t\}$, and $w_+(\cdot)$ is a standard,
n-dimensional {\it Wiener process} with the property that $w_+(t)-w_+(s)$ is independent of
$\{\xi(\tau);0\le \tau\le s\}$. Moreover, $\beta$ must satisfy the finite-energy condition
\begin{equation}\label{K2} E\left\{\int_{t_0}^{t_1}\beta(t)\cdot\beta(t)dt\right\}<\infty.
\end{equation}
In \cite{F}, F\"{o}llmer has shown that a finite-energy diffusion also admits a reverse-time
differential. Namely, there exists a measurable function $\gamma(t)$ of the future
$\{\xi(\tau);t\le \tau\le t_1\}$ called {\it backward drift}, and another Wiener process $w_-$
such that \begin{equation}\label{K3} \xi(t)-\xi(s)=\int_s^t\gamma(\tau)d\tau+\sigma
[w_-(t)-w_-(s)],\quad t_0\le s<t\le t_1. \end{equation}  Moreover, $\gamma$ satisfies
\begin{equation}\label{K4} E\left\{\int_{t_0}^{t_1}\gamma(t)\cdot\gamma(t)dt\right\}<\infty,
\end{equation} and $w_-(t)-w_-(s)$ is independent of $\{\xi(\tau);t\le \tau\le t_1\}$. Let us
agree that $dt$ always indicates a strictly positive variable. For any function $f$ defined on
$[t_0,t_1]$, let

$$d_+f(t):=f(t+dt)-f(t)$$ be the {\it forward increment} at time $t$, and $$d_-f(t)=f(t)-f(t-dt)$$
be the {\it backward increment} at time $t$. For a finite-energy diffusion, F\"{o}llmer has also
shown in \cite{F} that the forward and backward drifts may be obtained as Nelson's conditional
derivatives, namely \begin{equation}\label{D1}\beta(t)=\lim_{dt\searrow
0}E\left\{\frac{d_+\xi(t)}{dt}|\xi(\tau), t_0 \le \tau \le t\right\},\end{equation} and 
\begin{equation}\label{D2}\gamma(t)=\lim_{dt\searrow 0}E\left\{\frac{d_-\xi(t)}{dt}|\xi(\tau), t
\le  \tau \le t_1\right\},\end{equation} the limits being taken in $L^2_n(\Omega,{\cal B},P)$. It
was finally shown in \cite{F} that the one-time probability density $\rho(\cdot,t)$ of $\xi(t)$
(which exists for every $t>t_0$) is absolutely continuous on $\R^n$ and the following relation
holds a.s. $\forall t>0$ \begin{equation}\label{K4'} E\{\beta(t)-\gamma(t)|\xi(t)\} =
\sigma^2\nabla\log\rho(\xi(t),t). \end{equation}
Let  $\xi$ be a
finite-energy diffusion satisfying (\ref{K1}) and (\ref{K3}). Let $f:\R^n\times
[t_0,t_1] \rightarrow \R$ be twice continuously differentiable with respect to the spatial variable
and once with respect to time.
Then, we have the following change of variables formulas: 
\begin{eqnarray}
f(\xi(t),t)-f(\xi(s),s)=\int_s^t\left(\frac{\partial}{\partial
\tau}+\beta(\tau)\cdot\nabla+\frac{\sigma^2}{2}\Delta\right)f(\xi(\tau),\tau
)d\tau\\+\int_s^t\sigma\nabla f(\xi(\tau),\tau)\cdot
d_+w_+(\tau),\label{K5}\\f(\xi(t),t)-f(\xi(s),s)=
\int_s^t\left(\frac{\partial}{\partial
\tau}+\gamma(\tau)\cdot\nabla-\frac{\sigma^2}{2}\Delta\right)f(\xi(\tau),\tau)d\tau
\\+\int_s^t\sigma\nabla f(\xi(\tau),\tau)\cdot d_-w_-(\tau). \label{K6} \end{eqnarray} The
stochastic integrals appearing in (\ref{K5}) and (\ref{K6}) are a (forward) Ito integral and a
backward Ito integral, respectively, see \cite {N3} for the details. Let us introduce the {\it
current drift} $v(t):=(\beta(t)+\gamma(t))/2$ and the {\it osmotic drift}
$u(t):=(\beta(t)-\gamma(t))/2$. Notice that, when $\sigma$ tends to zero, $v$ tends to
$\dot{\xi},$ and $u$ tends to zero. The semi-sum and the semi-difference of (\ref{K5}) and
(\ref{K6}) give two more useful formulas: \begin{eqnarray}\nonumber
&&f(\xi(t),t)-f(\xi(s),s)=\int_s^t\left(\frac{\partial}{\partial
\tau}+v(\tau)\cdot\nabla\right)f(\xi(\tau),\tau)d\tau\\&&+\frac{\sigma}{2}\left[\int_s^t\nabla
f(\xi(\tau),\tau)\cdot d_+w_+ +\int_s^t\nabla f(\xi(\tau),\tau)\cdot
d_-w_-\right],\label{K7}\end{eqnarray} \begin{eqnarray} &&0=\nonumber\int_s^t\left(
u(\tau)\cdot\nabla+\frac{\sigma^2}{2}\Delta\right)f(\xi(\tau),\tau)d\tau\\&&+\frac{\sigma}{2}\left[\int_s^t\nabla
f(\xi(\tau),\tau)\cdot d_+w_+-\int_s^t\nabla f(\xi(\tau),\tau)\cdot d_-w_-\right]. \label{K8}
\end{eqnarray} Specializing (\ref{K7}) and (\ref{K8}) to $f(x,t)=x$, we get
\begin{eqnarray}\label{K9}\xi(t)-\xi(s)&=&\int_s^tv(\tau)d\tau
+\frac{\sigma}{2}\left[w_+(t)-w_+(s)+w_-(t)-w_-(s)\right],\\0&=&\int_s^tu(\tau)d\tau+
\frac{\sigma}{2}\left[w_+(t)-w_+(s)-w_-(t)+w_-(s)\right]\label{K10} \end{eqnarray}

The finite-energy diffusion $\xi(\cdot)$ is called {\it Markovian} if there exist two measurable
functions $b_+(\cdot,\cdot)$ and $b_-(\cdot,\cdot)$ such that $\beta(t)=b_+(\xi(t),t)$ a.s. and
$\gamma(t)=b_-(\xi(t),t)$ a.s., for all $t$ in $[t_0,t_1]$. The duality relation (\ref{K4'}) now
reads \begin{equation}\label{M1} b_+(\xi(t),t)-b_-(\xi(t),t) = \sigma^2\nabla\log\rho(\xi(t),t).
\end{equation} This immediately gives the {\it osmotic equation} \begin{equation}\label{M2}
u(x,t)=\frac{\sigma^2}{2}\nabla\log\rho(x,t), \end{equation} where
$u(x,t):=(b_+(x,t)-b_-(x,t))/2$. The probability density $\rho(\cdot,\cdot)$ of $\xi(t)$ satisfies
(at least weakly) the {\it Fokker-Planck equation} $$ \frac{\partial{\rho}}{\partial{t}} + \nabla
\cdot (b_+\rho) = \frac{\sigma^2}{2}\Delta\rho. $$  The latter can also be rewritten, in view of
(\ref{M1}), as the {\it equation of continuity} of hydrodynamics \begin{equation}\label{M3}
\frac{\partial{\rho}}{\partial{t}} + \nabla \cdot (v\rho) = 0, \end{equation} where
$v(x,t):=(b_+(x,t)+b_-(x,t))/2$.

\section{A time-symmetric kinematics for diffusion processes}

We recall here the basic facts from the time-symmetric kinematics developed in 
\cite{P1,P4}. Let us multiply
(\ref{K10}) by $-i$, and add it to (\ref{K9}). We get 
\begin{eqnarray}\nonumber
&&\xi(t)-\xi(s)=\int_s^t[v(\tau)-iu(\tau)]d\tau\\&&
+\frac{\sigma}{2}\left[(1-i)(w_+(t)-w_+(s))+(1+i)(w_-(t)-w_-(s))\right]. \label{K11}\end{eqnarray}  We
call $v_q(t):=v(t)-iu(t)$ the {\it quantum drift}, and
\begin{equation}\label{K12}w_q(t):=\frac{1-i}{2}w_+(t)+\frac{1+i}{2}w_-(t)\end{equation} the
{\it quantum noise}. Hence, we can rewrite (\ref{K11}) as \begin{equation}\label{K13}
\xi(t)-\xi(s)=\int_s^tv_q(\tau)d\tau+\sigma[w_q(t)-w_q(s)]. \end{equation}
 At first sight, this decomposition of the {\it real-valued} increments of $\xi$ into the sum of
two {\it complex} quantities might look somewhat odd. Nevertheless, this representation enjoys
several important properties. 
\begin{enumerate} \item When $\sigma^2$ tends to zero, $v-iu$ tends
to $\dot{\xi}$. 
\item The quantum drift $v_q(t)$ contains at each time $t$ precisely the same
information as the pair $(v(t),u(t))$ (or, equivalently, the pair $(\beta(t),\gamma(t))$. 
\item The representation (\ref{K13}), differently from
(\ref{K1}) and (\ref{K3}) enjoys an important symmetry with respect to time. Indeed, under time
reversal, (\ref{K13}) transforms into \begin{equation}\label{K14}
\xi(t)-\xi(s)=\int_s^t\overline{v_q(\tau)}d\tau+\sigma[\overline{w_q(t)-w_q(s)}],
\end{equation} where overbar indicates conjugation, see \cite[p.145]{P2}.  \end{enumerate}  The
representation (\ref{K13}) has proven to be crucial in order to develop a Lagrangian and
Hamiltonian dynamics formalism in the context of Nelson's stochastic mechanics, see
\cite{P1}-\cite{P3}. In particular, to develop the second form of Hamilton's principle, the
key tool has been a change of variables formula related to representation (\ref{K13}). In
order to recall such a formula, we need first to define stochastic integrals with respect to
the quantum noise $w_q$. Let us denote by
$d_bf(t):=\frac{1-i}{2}d_+f(t)+\frac{1+i}{2}d_-f(t)$ the  {\it bilateral increment} of $f$ at time
$t$. Then, from (\ref{K12}) and (\ref{K10}), we get 
$$d_+w_q(t)=\frac{1+i}{\sigma}u(x(t),t)dt+ d_+w_+ +o(dt),$$
$$d_-w_q(t)=\frac{-1+i}{\sigma}u(x(t),t)dt+ d_+w_- +o(dt).$$ These in turn give immediately
\begin{equation}\label{QN}d_bw_q(t):=\frac{1-i}{2}d_+w_+(t)+\frac{1+i}{2}d_-w_-(t) +o(dt).
\end{equation} Let
$f(x,t)$ be a measurable, $\C^n$-valued function such that
$$P\left\{\omega:\int_0^Tf(\xi(t),t)\cdot\overline{f(\xi(t),t)}dt <\infty\right\}=1. $$ In
view of (\ref{QN}), we define  
$$\int_s^tf(\xi(\tau),\tau)\cdot
d_bw_q(\tau):=\frac{1-i}{2}\int_s^tf(\xi(\tau),\tau)\cdot
d_+w_+(\tau)+\frac{1+i}{2}\int_s^tf(\xi(\tau),\tau)\cdot d_-w_-(\tau).$$
Thus, integration with respect to the bilateral increments of $w_q$ is defined through a linear
combination with complex coefficients of a forward and a backward Ito integral. Let $f(x,t)$ be
a complex-valued function with real and imaginary parts of class $C^{2,1}$. Then, multiplying
(\ref{K8}) by $-i$, and then adding it to (\ref{K7}), we get the change of variables formula
\begin{eqnarray}\nonumber f(\xi(t),t)-f(\xi(s),s)&=&\int_s^t\left(\frac{\partial}{\partial
\tau}+v_q(\tau)\cdot\nabla-\frac{i\sigma^2}{2}\Delta\right)f(\xi(\tau),\tau)d\tau
\nonumber\\&+&\int_s^t\sigma\nabla
f(\xi(\tau),\tau)\cdot d_bw_q(\tau).\label{K16} \end{eqnarray} It is important to understand
that this formula, and in particular the coefficient of the Laplacian term, follows from basic
probabilistic arguments.

\section{The quantum Hamilton principle} 
Stochastic mechanics may be based, since the fundamental paper by Guerra and Morato
\cite{GM}, on stochastic variational principles of hydrodynamic type. Other versions of the
variational principle have been proposed in \cite{N2,N4}, and in \cite{P1}. We outline
here the quantum Hamilton principle of \cite{P1}, since it employs the time-symmetric kinematics
of Section 3 that we shall need to derive the wavefunction collapse.

Let
${\cal X}_{\rho_1}$ denote the family of all finite-energy, $\R^n$-valued diffusions on
$[t_0,t_1]$ with diffusion coefficient
$I_n\frac{\hbar}{m}$, and having marginal probability density $\rho_1$ at time $t_1$. Let $\cal V$
denote the family of finite-energy, $C^n$ - valued stochastic processes on $[t_0,t_1]$. Let
$L(x,v) := \frac{1}{2} m v\cdot v - V(x)$ be defined on $R^n\times C^n$. Also let $S_0$ be a
complex-valued function on $R^n$. Consider the problem of extremizing on $(x,v_q)\in({\cal
X}_{\rho_1}\times\cal V)$ \begin{equation}\label{H5} E\left\{\int_{t_0}^{t_1} L(x(t),v_{q}(t))
\,dt + S_{0}(x(t_{0})\right\} \end{equation} subject to the constraint that
\begin{equation}\label{H5'} x \;{\rm has\;quantum\;drift\;(velocity)}\; v_q. \end{equation}
Notice that the quadratic term in the Lagrangian may be rewritten in terms of the forward and
backward drifts as follows \begin{eqnarray}\nonumber
\frac{m}{2}v_q(t)\cdot
v_q(t)=\frac{m}{2}[(\frac{1-i}{2}\beta(t)+\frac{1+i}{2}\gamma(t)]\cdot 
[(\frac{1-i}{2}\beta(t)+\frac{1+i}{2}\gamma(t)]=\\\frac{-im}{4}
[\beta(t)\cdot\beta(t)+2i\beta(t)\cdot\gamma(t)-\gamma(t)\cdot\gamma(t)]
=\frac{-im}{4}[(\beta(t)+i\gamma(t))\cdot(\beta(t)+i\gamma(t))]
\label{HHH} \end{eqnarray}
In
\cite[Section VIII]{P1}, the following result was established. 
\begin{theorem} Suppose that
$S_q(x,t)$ of class $C^{2,1}$ solves on $[t_0,t_1]$ the initial value problem
\begin{eqnarray}\label{H6} &&\frac{\partial{S_q}}{\partial{t}} + \frac{1}{2m}\nabla {S_q} \cdot
\nabla{S_q} + V(x) - \frac{i\hbar}{2m}\Delta{S_q} = 0,\\&&S_q(x,t_0)=S_0(x),\label{H6'}
\end{eqnarray} and satisfies the technical condition \begin{equation}\label{H6"}
E\left\{\int_{t_0}^{t_1}\nabla S_q(x(t),t)\cdot\overline{\nabla
S_q(x(t),t)}\;dt\right\}<\infty,\quad \forall x\in {\cal X}_{\rho_1}. \end{equation} Then, any
$x\in {\cal X}_{\rho_1}$ having quantum drift $\frac{1}{m}\nabla S(x(t),t)$ solves the
extremization problem.
\end{theorem}
A crucial role in the proof is played by the change of variables formula (\ref{K16}) that here
reads
\begin{eqnarray}\nonumber f(\xi(t),t)-f(\xi(s),s)&=&\int_s^t\left(\frac{\partial}{\partial
\tau}+v_q(\tau)\cdot\nabla-\frac{i\hbar}{2m}\Delta\right)f(\xi(\tau),\tau)d\tau
\nonumber\\&+&\int_s^t\sqrt{\frac{\hbar}{m}}\;\;\nabla
f(\xi(\tau),\tau)\cdot d_bw_q(\tau).\label{K160} \end{eqnarray}
Existence of a solution for the apparently complicated nonlinear, complex Cauchy problem
(\ref{H6})-(\ref{H6'}) is dealt with as follows. Let $\{\psi(x,t); t_0\le t\le t_1\}$ be the
solution of the {\it Schr\"{o}dinger equation} 
\begin{equation}\label{H7} \frac{\partial{\psi}}{\partial{t}} = \frac{i\hbar}{2m}\Delta\psi -
\frac{i}{\hbar}V(x)\psi, \end{equation} with initial condition
$\psi_{0}(x):=\exp{\frac{i}{\hbar}S_{0}(x)}$. If $\psi(x,t)$ never vanishes on $\R^n\times
[t_0,t_1]$, and satisfies the condition \begin{equation} E\left\{\int_{t_0}^{t_1}\nabla
\log\psi(x(t),t)\cdot\overline{\nabla \log\psi(x(t),t)}\;dt\right\}<\infty,\quad \forall x\in
{\cal X}_{\rho_1}, \end{equation} then $S_q(x,t):=\frac{\hbar}{i}\log\psi(x,t)$ satisfies
(\ref{H6})-(\ref{H6'}) and (\ref{H6"}). If, moreover,  $\psi_{0}(x)$ has $L^2$ norm $1$, and the
terminal density satisfies $\rho_1(x,t)=|\psi(x,t_1)|^2$, then there does exist a Markov diffusion
having the required quantum drift, namely the {\it Nelson process} associated to $\{\psi(x,t);
t_0\le t\le t_1\}$, and Born's relation $\rho(x,t) = |\psi(x,t)|^2$  holds, see \cite{P1} for the
details. The construction of the Nelson process corresponding to $\psi(x,t)$ in the case where
$\psi(x,t)$ vanishes requires considerable care. It is discussed in \cite{C}, \cite [Chapter
IV]{BCZ}, and references therein.

\section{Measurement in nonequilibrium thermodynamics}
In this section, we
discuss measurement for nonequilibrium thermodynamical systems. This serves as an
introduction to measurement in
stochastic mechanics to be discussed in the following section. Consider an open
thermodynamical system whose macroscopic evolution is modeled by an n-dimensional
Markov diffusion process
$\{x(t);t_0\le t\}$ with forward Ito differential
$$d_+x(t)=b_+(x(t))dt+\sigma d_+w_+.$$ Let $\rho(x,t)$ denote the probability density of
$x(t)$ satisfying the Fokker-Planck equation
\begin{equation}\label{HH0}\frac{\partial{\rho}}{\partial{t}} + \nabla
\cdot (b_+\rho) = \frac{\sigma^2}{2}\Delta\rho. 
\end{equation}
The {\it equilibrium state} is given by 
the Maxwell-Boltzmann distribution
law  $$\bar{\rho}(x) = C exp[ - \frac{H(x)}{kT} ],$$ where $H$ is the
Hamiltonian function, and we have the relation
$$b_+(x)=-\frac{\sigma^2}{2kT}\nabla H(x),
$$
where $k$ is Boltzmann's constant and $T$ is the absolute temperature. Suppose that at time $t_1$ a
measurement is made that yields the new probability density $\tilde{\rho}(x,t_1)$. Let ${\cal
X}_{\tilde{\rho}(t_1)}$ denote the class of finite-energy diffusions on $[t_1,t_2]$ with
diffusion coefficient
$\sigma^2$ and having marginal $\tilde{\rho}(x,t_1)$ at time $t_1$. Let us pose the following
question: Among all processes in  ${\cal X}_{\tilde{\rho}(t_1)}$, which one should we use
to model the macroscopic evolution of the system from $t_1$ up to $t_2$? Everybody agrees that we
should employ the stochastic process $\{\tilde{x}(t);t_1\le t\le t_2\}$ that has the same forward drift
field $b_+(x)$ of the ``reference" process $x$. This is supported by the observation that the new
process must have the same equilibrium distribution of the previous one. Let
us show that the new process  $\{\tilde{x}(t);t_1\le t\le t_2\}$ may be obtained 
as solution of a variational
problem. Assume that the Kullback-Leibler pseudo-distance between  $\tilde{\rho}(t_1)$ and 
$\rho(t_1)$ is finite, namely
$$H(\tilde{\rho}(t_1),\rho(t_1)):=E\left\{\log\frac{\tilde{\rho}(\tilde{x}(t_1),t_1)}
{\rho(\tilde{x}(t_1),t_1)}\right\}=\int_{\R^n}\log\frac{\tilde{\rho}(\tilde{x},t_1)}
{\rho(\tilde{x},t_1)}\tilde{\rho}(x,t_1)dx <\infty.$$
Let $\D_{\tilde{\rho}(t_1)}$ denote the class of probability measures on $\Omega=C([t_1,t_2])$
that are equivalent to the measure $P$ induced by the reference process $\{x(t);t_1\le t\le t_2\}$. For
$Q\in\D_{\tilde{\rho}(t_1)}$, let  $$H(Q,P)=E_Q[\log\frac{dQ}{dP}]$$ denote the {\it relative
entropy } of $Q$ with respect to $P$. It then follows from Girsanov's theorem that \cite{F,F2}
$$H(Q,P)=H(\tilde{\rho}(t_1),\rho(t_1))+E_Q\left[\int_{t_1}^{t_2}\frac{1}{2\sigma^2}
[b_+(\tilde{x}(t))-\beta^Q(t)]\cdot [b_+(\tilde{x}(t))-\beta^Q(t)]dt\right].
$$
Since $H(\tilde{\rho}(t_1),\rho(t_1))$ is constant over $\D_{\tilde{\rho}(t_1)}$, it trivially
follows that the probability measure $\tilde{Q}$ corresponding to the process $\tilde{x}$ having
forward drift $b_+$ minimizes $H(Q,P)$ over $\D_{\tilde{\rho}(t_1)}$.
This problem may be interpreted as a problem of large deviation of the empirical distribution according to
Schr\"{o}dinger's original motivation \cite{S,F2}.  We consider now an apparently different variational problem
that has the same solution as the previous one. We do so, because it is this second form which, in a suitably
modified form, applies to the quantum case. Let ${\cal X}_{\tilde{\rho}_2}$ denote the family of finite-energy
diffusions on $[t_1,t_2]$ with diffusion coefficient $\sigma^2$ and having marginal density $\tilde{\rho}_2$ at
time $t_2$. Consider the problem of minimizing with respect to the pair $(\tilde{x},\gamma)$ the functional
$$E\left\{\int_{t_1}^{t_2}\frac{1}{2\sigma^2}[b_-(\tilde{x}(t))-\gamma(t)]\cdot
[b_-(\tilde{x}(t))-\gamma(t)]dt-\log\frac{\tilde{\rho}(\tilde{x}(t_1),t_1)}
{\rho(\tilde{x}(t_1),t_1)}\right\} $$ subject to the constraint that $\gamma$ be the backward
drift of $\tilde{x}$ on $[t_1,t_2]$. This problem is a variant of the one first considered
and solved in \cite[Theorem 2]{GP}. The connection between the two variational problems, and their relation
to the theory of Schr\"{o}dinger processes and bridges, has been thoroughly investigated in \cite{PW}. 
In order to solve this problem, rather than reproducing the arguments in \cite{GP,PW}, we take the opportunity
to introduce the variational method based on nonlinear Lagrange functionals, \cite{KP3}. This method
permits to solve also the more complicated quantum case. Suppose that we wish to minimize
$J:Y\rightarrow \bar{\R}$, where $\bar{\R}$ denotes the extended reals, over the nonempty subset $S$
of $Y$.  \begin{lemma}\label{L}{\em (Lagrange Lemma)} Let $\Lambda
:Y\rightarrow
\bar{\R}$ and let $y_0\in S$ minimize $J+\Lambda$ over $Y$. Assume that
$\Lambda(\cdot)$ is {\em finite} and {\em constant} over $S$. Then
$y_0$ minimizes $J$ over $S$.
\end{lemma} 
\proof For any
$y\in S$, we have
$J(y_0)+\Lambda(y_0) \le J(y)+\Lambda(y) = J(y)+\Lambda(y_0)$.
Hence $J(y_0)\le J(y)$. \qed  
\noindent
A functional $\Lambda$ which is constant and finite on $S$ is called a {\it Lagrange functional}.
Obviously, a similar result holds if the problem is an extremization problem. Let us apply this simple
idea to the above problem. Let
$\varphi(x,t)$ be a real-valued function of class $C^{2,1}$ defined on $\R^n\times [t_1,t_2]$, and
satisfying the technical condition \begin{equation}\label{HH}E\left\{\int_{t_1}^{t_2}\nabla
\varphi(x(t),t)\cdot\nabla \varphi(x(t),t)\;dt\right\}<\infty,\quad \forall x\in
{\cal X}_{\tilde{\rho}_2}.  
\end{equation}
Corresponding to such a $\varphi$, we introduce the functional
\begin{eqnarray}\nonumber&&\Lambda^{\varphi}(\tilde{x},\gamma):=E\left\{
\varphi(\tilde{x}(t_2),t_2)-\varphi(\tilde{x}(t_1),t_1)\right.\\&&\nonumber\left.
+\int_{t_1}^{t_2}\left[-\frac{\partial{\varphi}}{\partial{t}}(\tilde{x}(t),t) -
\gamma(t)\cdot \nabla \varphi(\tilde{x}(t),t)+\frac{\sigma^2}{2}\Delta
\varphi(\tilde{x}(t),t)\right]dt\right\}. 
\end{eqnarray}
In view of (\ref{K6}) and (\ref{HH}), we have that
$\Lambda^{\varphi}(\tilde{x},\gamma)=0$
whenever the pair $(\tilde{x},\gamma)$ satisfies the constraint since the stochastic integral
has zero expectation. Thus, it is a  {\it Lagrange functional} for the problem. Consider next
the {\it unconstrained} minimization of the functional $J+\Lambda^{\varphi}$. For a fixed
$\tilde{x}\in {\cal X}_{\tilde{\rho}_2}$, and a fixed time $t\in [t_1,t_2]$, we consider the
{\it pointwise} minimization of the integrand of $J+\Lambda^{\varphi}$ with respect to $\gamma$
$$ {\rm minimize}_{\gamma\in
R^n}\{\frac{1}{2\sigma^2}(b_-(\tilde{x}(t),t)-\gamma)\cdot(b_-(\tilde{x}(t),t)-\gamma)
- \gamma\cdot \nabla \varphi(\tilde{x}(t),t)\}$$
We get
\begin{equation}\label{HH1}
\gamma^o(\tilde{x})(t)=b_-(\tilde{x}(t),t)+\sigma^2\nabla\varphi(\tilde{x}(t),t).
\end{equation}
Substituting back expression (\ref{HH1}) into $J+\Lambda^{\varphi}$, we get  the following
functional of $\tilde{x}$
\begin{eqnarray}\nonumber(J+\Lambda^{\varphi})(\tilde{x},\gamma^o(\tilde{x})):=E\left\{
\varphi(\tilde{x}(t_2),t_2)-\varphi(\tilde{x}(t_1),t_1)-\log\frac{\tilde{\rho}(\tilde{x}(t_1),t_1)}
{\rho(\tilde{x}(t_1),t_1)}+\right.\\\nonumber\left.
\int_{t_1}^{t_2}\left[-\frac{\sigma^2}{2}\nabla\varphi(\tilde{x}(t),t)\cdot\nabla\varphi(\tilde{x}(t),t)-\frac{\partial{\varphi}}{\partial{t}}(\tilde{x}(t),t)
\nonumber\right.\right.\\\left.\left.- b_-(\tilde{x}(t),t)\cdot \nabla
\varphi(\tilde{x}(t),t)+\frac{\sigma^2}{2}\Delta \varphi(\tilde{x}(t),t)\right]dt\right\}. 
\end{eqnarray}
Next, we seek to find a function $\varphi$ such that the functional 
$(J+\Lambda^{\varphi})(\tilde{x},\gamma^o(\tilde{x}))$
becomes constant over ${\cal X}_{\tilde{\rho}_2}$. Suppose $\varphi$ solves on $[t_1,t_2]$ the
initial value problem \begin{eqnarray}\label{HH2}\frac{\partial{\varphi}}{\partial{t}}
+ b_-(x,t)\cdot \nabla \varphi(x,t)-\frac{\sigma^2}{2}\Delta
\varphi(x,t)=-\frac{\sigma^2}{2}\nabla\varphi(x,t)\cdot\nabla\varphi(x,t),\\\label{HH3}
\varphi(x,t_1)=-\log\frac{\tilde{\rho}(x,t_1)}{\rho(x,t_1)}.
\end{eqnarray}
Then  $(J+\Lambda^{\varphi})(\tilde{x},\gamma^o(x))=E\{\varphi(\tilde{x}(t_2),t_2)\}$ is
constant  over ${\cal X}_{\tilde{\rho}_2}$ since such processes have the same marginal
density at time $t_2$. Hence, any $x\in {\cal X}_{\tilde{\rho}_2}$ solves the unconstrained
minimization of $J+\Lambda^{\varphi}$.  To solve the
original constrained problem, we need to find  $\tilde{x}\in{\cal X}_{\tilde{\rho}_2}$
that has backward drift given by (\ref{HH1}). In order to do that, we first proceed to find the
solution of (\ref{HH2})-(\ref{HH3}). Define
$\tilde{\rho}(x,t):=\exp [-\varphi(x,t)] \rho(x,t)$. Then, if $\varphi$ satisfies (\ref{HH2}),
using the Fokker-Plank equation satisfied by $\rho$, we get 
\begin{eqnarray}\nonumber&&\frac{\partial\tilde{\rho}}{\partial t}=\exp [-\varphi]
\left(-\frac{\partial \varphi}{\partial t}\rho+ \frac{\partial \rho}{\partial
t}\right)=\\&&\nonumber
\left(b_-\cdot\nabla\varphi-\frac{\sigma^2}{2}\Delta\varphi+\frac{\sigma^2}{2}\nabla\varphi\cdot\nabla\varphi\right)\tilde{\rho}
-\exp[-\varphi]\nabla\cdot(b_+\rho)+\exp[-\varphi]\frac{\sigma^2}{2}\Delta\rho=\\&&
\frac{\sigma^2}{2}\Delta\tilde{\rho}+b_+\cdot\nabla\varphi\tilde{\rho}
-\exp[-\varphi]\nabla\rho\cdot b_+-\exp[-\varphi]\rho\nabla\cdot
b_+=-\nabla\cdot(\tilde{\rho}b_+)+\frac{\sigma^2}{2}\Delta\tilde{\rho}\nonumber.
\end{eqnarray}
We conclude that if $\tilde{\rho}$ is the solution of 
the Fokker-Planck equation (\ref{HH0}) on $[t_1,t_2]$ with initial condition at time $t_1$
given by $\tilde{\rho}(x,t_1)$, then
$\varphi:=-\log\frac{\tilde{\rho}}{\rho}$ solves the initial value problem
(\ref{HH2})-(\ref{HH3}). Thus, we have the following result. 

\begin{theorem} Let $\tilde{\rho}$ be the solution of the Fokker-Planck equation (\ref{HH0}) on
$[t_1,t_2]$ with initial condition given by
$\tilde{\rho}(x,t_1)$. Then 
$\varphi:=-\log\frac{\tilde{\rho}}{\rho}$ solves the initial value problem
(\ref{HH2})-(\ref{HH3}). Suppose that $\varphi$
satisfies (\ref{HH}), and that $\tilde{\rho}_2(x)=\tilde{\rho}(x,t_2)$. Then the stochastic
process $\tilde{x}\in {\cal X}_{\tilde{\rho}_2}$ having backward drift field
$\tilde{b}_-(x,t)=b_-(x,t)-\sigma^2\nabla\log\frac{\tilde{\rho}}{\rho}(x,t)=b_+
-\sigma^2\nabla\log\tilde{\rho}(x,t)$ solves the constrained minimization problem. 
\end{theorem}
\noindent
In view of (\ref{M1}), we see that the solution process has forward drift $b_+(\cdot)$, and
therefore coincides with the solution of the previous variational problem. 
Consider the same problem on the interval $[t_1,t_3]$, where $t_3>t_2$. If we impose the density
$\tilde{\rho}(x,t_3)$ at the final time, the solution process coincides with the previous
solution process up to time $t_2$. This may be viewed as a form of coherence with respect to the
terminal time. It is also important to observe that the new process  $\{\tilde{x}(t);t_1\le t\le
t_2\}$ has the same forward drift of the reference process $\{x(t);t_1\le t\le t_2\}$, but a {\it
different backward drift}. Hence, while the forward transition probabilities have been preserved,
{\it the reverse-time transition probabilities have changed}. Thus, we see that it is impossible,
even in principle, to estimate the reverse-time transition probabilities by repeated measurement.
In \cite{N4,G94}, Nelson and Guerra regard as a serious drawback of stochastic mechanics the fact
that transition probabilities of the Nelson process are not open to experimental verification if
we accept that transition probabilities are associated to a definite quantum state. We shall come
back to this crucial point in the next section.

\section{A stochastic derivation of wave function collapse}
In Section 4, we have seen that the Schr\"{o}dinger equation is obtained  through a
simple exponential transformation from the Hamilton-Jacobi equation (\ref{H6}) of an appropriate
stochastic variational principle. Suppose now that a position measurement of the quantum system
is made at time $t_1$, and we ask: What should be the new stochastic process on $[t_1,t_2]$?
First of all, we consider the situation without measurement up to time $t_2$. In this case, the
variational principle of  Section 4 would have as solution the Nelson process $\{x(t); t_0\le
t\le t_2\}$ extended up to time $t_2$ with quantum drift
$v_q(t)=\frac{\hbar}{im}\nabla\log\psi(x(t),t)$,  where $\{\psi(x,t):t_0\le t \le t_2\}$ is the
solution of  the Schr\"{o}dinger equation (\ref{H7}). The Nelson process  $\{x(t); t_1\le
t\le t_2\}$ will play the role of a ``reference process". Suppose that the measurement at time
$t_1$ yields the new probability density $\tilde{\rho}(x,t_1)$. For instance, if we assume
that the measurement at time $t_1$ only gives the information that
$x$ lies in a certain subset
$D$ of the configuration space of the system, the density $\tilde{\rho}(x,t_1)$
just after the measurement is given, according to Bayes' theorem, by
$$\tilde{\rho}(x,t_1)=\frac{\chi_D(x)\rho(x,t_1)}{\int_D \rho(x',t_1)dx'}\;,
$$
where $\rho(x,t_1)$ is the probability density of the Nelson reference process right before the
measurement is made. We need now to find an appropriate variational mechanism that, employing
the Nelson reference process and the probability density $\tilde{\rho}(x,t_1)$, produces the
new process
$\{\tilde{x}(t); t_1\le t\le t_2\}$. It is apparent that the variational mechanism of the
previous section is not suitable here. Indeed, as observed before, that mechanism preserves
completely the {\it forward} drift and transition probabilities, but changes, possibly in a
dramatic way, the backward drift and transition probabilities. This is not acceptable in
stochastic mechanics, were forward and backward drifts and transition probabilities {\it must
always be granted the same status}. In other words, the time-reversibility of the theory must be
reflected also by the theory of measurement. On the other hand, preserving both drifts, or equivalently
both transition probabilities, amounts to preserving the process  $\{x(t); t_0\le t\le t_2\}$, which is
impossible since the probability density at time $t_1$ has changed. Thus, we need to find a
variational mechanism that {\it changes both drifts as little as possible, given the new density at
time} $t_1$. In should be apparent that, at this point, the time-symmetric kinematics of Section 3 is
called for. Given that kinematics, and by analogy with the variational principle of the previous
section, we are then led to the following formulation.
 
In the notation of Section 4, we consider the problem of extremizing on
$(\tilde{x},\tilde{v}_q)\in({\cal X}_{\tilde{\rho}_2}\times\cal V)$ the functional 
\begin{equation}J(\tilde{x},\tilde{v_q}):=E\left\{\int_{t_1}^{t_2}
\frac{mi}{2\hbar}(v_q(\tilde{x}(t),t)-\tilde{v}_q(t))\cdot(v_q(\tilde{x}(t),t)-\tilde{v}_q(t))
\,dt +
\frac{1}{2}\log\frac{\tilde{\rho}(\tilde{x}(t_1),t_1)}{\rho(\tilde{x}(t_1),t_1)}\right\}
\label{H8}\end{equation}   subject to the
constraint that \begin{equation}\label{H9} \tilde{x} \;{\rm has\;quantum\;drift\;(velocity)}\;
\tilde{v}_q. \end{equation}
Here  $v_q(x,t)=\frac{\hbar}{im}\nabla\log\psi(x,t)$ is the quantum drift field of the
Nelson reference process, and ${\cal X}_{\tilde{\rho}_2}$ is the family of all finite-energy,
$\R^n$-valued diffusions on $[t_1,t_2]$ with diffusion coefficient $I_n\frac{\hbar}{m}$, and
having probability density $\tilde{\rho}_2$ at time $t_2$. The structure of the functional is
quite similar to the one of the previous section. Here, $\frac{\hbar}{mi}$ replaces $\sigma^2$ in
view of formula (\ref{K160}). The $\frac{1}{2}$ in the boundary term is justified by the
following relation, see (\ref{HHH}), \begin{eqnarray}\nonumber
&&\frac{mi}{2\hbar}(v_q(x,t)-\tilde{v}_q(t))\cdot(v_q(x,t)-\tilde{v}_q(t))
=\\&&\frac{m}{4\hbar}\left[(b_+(x,t)-\tilde{b}_+
(t))+i(b_-(x,t)-\tilde{b}_-(t))\right]\cdot\left[(b_+(x,t)-\tilde{b}_+
(t))+i(b_-(x,t)-\tilde{b}_-(t))\right]\nonumber
\end{eqnarray}
which shows that a $\frac{1}{4}$ appears in the right-hand side.
To solve this variational problem, we employ the same strategy as in the previous section. Let
$\varphi(x,t)$ be a complex-valued function of class $C^{2,1}$ defined on $\R^n\times
[t_1,t_2]$, and satisfying the technical condition
\begin{equation}\label{H10}E\left\{\int_{t_1}^{t_2}\nabla
\varphi(x(t),t)\cdot\overline{\nabla \varphi(x(t),t)}\;dt\right\}<\infty,\quad \forall x\in
{\cal X}_{\rho_2}.  
\end{equation}
Corresponding to such a $\varphi$, we introduce the functional
\begin{eqnarray}\nonumber&&\Lambda^{\varphi}(\tilde{x},\tilde{v_q}):=E\left\{
\varphi(\tilde{x}(t_2),t_2)-\varphi(\tilde{x}(t_1),t_1)+\right.\\&&\nonumber\left.
\int_{t_1}^{t_2}\left[-\frac{\partial{\varphi}}{\partial{t}}(\tilde{x}(t),t) -
\tilde{v}_q(t)\cdot \nabla \varphi(\tilde{x}(t),t)+\frac{i\hbar}{2m}\Delta
\varphi(\tilde{x}(t),t)\right]dt\right\}. 
\end{eqnarray}
In view of (\ref{K16}), and of property (\ref{H10}), we see that
$\Lambda^{\varphi}(\tilde{x},\tilde{v_q})=0$
whenever the pair $(\tilde{x},\tilde{v_q})$ satisfies the constraint. Thus, it is a 
{\it Lagrange functional} for the problem. Consider next the {\it unconstrained} extremization
of the functional $J+\Lambda^{\varphi}$. For a fixed $\tilde{x}\in {\cal X}_{\tilde{\rho}_2}$, and
a fixed time $t\in [t_1,t_2]$, we consider the {\it pointwise} extremization of the integrand of
$J+\Lambda^{\varphi}$ with respect to $\tilde{v}_q$
$$ {\rm extremize}_{\tilde{v}\in
C^n}\{\frac{mi}{2\hbar}(v_q(\tilde{x}(t),t)-\tilde{v})\cdot(v_q(\tilde{x}(t),t)-\tilde{v})
- \tilde{v}\cdot \nabla \varphi(\tilde{x}(t),t)\}$$
We get
\begin{equation}\label{H11}
\tilde{v}_q^o(\tilde{x})(t)=v_q(\tilde{x}(t),t)+\frac{\hbar}{mi}\nabla\varphi(\tilde{x}(t),t).
\end{equation}
Substituting back expression (\ref{H11}) into $J+\Lambda^{\varphi}$, we get  the following
functional of $\tilde{x}$
\begin{eqnarray}\nonumber(J+\Lambda^{\varphi})(\tilde{x},\tilde{v}_q^o(x)):=E\left\{
\varphi(\tilde{x}(t_2),t_2)-\varphi(\tilde{x}(t_1),t_1)+\right.\\\nonumber\left.
\int_{t_1}^{t_2}\left[\frac{i\hbar}{2m}\nabla\varphi(\tilde{x}(t),t)\cdot\nabla\varphi(\tilde{x}(t),t)-\frac{\partial{\varphi}}{\partial{t}}(\tilde{x}(t),t)
\nonumber\right.\right.\\\left.\left.- v_q(\tilde{x}(t),t)\cdot \nabla
\varphi(\tilde{x}(t),t)+\frac{i\hbar}{2m}\Delta \varphi(\tilde{x}(t),t)\right]dt\right\}. 
\end{eqnarray}
We seek next to choose the function $\varphi$ so that the functional $(J+\Lambda^{\varphi})(\tilde{x},\tilde{v}_q^o(x))$
becomes constant over ${\cal X}_{\tilde{\rho}_2}$. Suppose $\varphi$ solves on $[t_1,t_2]$ the
initial value problem 
\begin{eqnarray}\label{H12}\frac{\partial{\varphi}}{\partial{t}}
+ v_q(x,t)\cdot \nabla \varphi(x,t)-\frac{i\hbar}{2m}\Delta
\varphi(x,t)=\frac{i\hbar}{2m}\nabla\varphi(x,t)\cdot\nabla\varphi(x,t),\\\label{H13}
\varphi(x,t_1)=\frac{1}{2}\log\frac{\tilde{\rho}(x,t_1)}{\rho(x,t_1)}.
\end{eqnarray}
Then  $(J+\Lambda^{\varphi})(\tilde{x},\tilde{v}_q^o(x))=E\{\varphi(\tilde{x}(t_2),t_2)\}$ is constant 
over ${\cal X}_{\tilde{\rho}_2}$ since such processes have have the same marginal density at time
$t_2$. Hence, any $x\in {\cal X}_{\tilde{\rho}_2}$ solves the unconstrained extremization of
$J+\Lambda^{\varphi}$. To solve the original constrained extremization problem, we need to find 
$\tilde{x}\in{\cal X}_{\tilde{\rho}_2}$ that has quantum drift given by (\ref{H11}). In order to
do that, we first proceed to find the solution of (\ref{H12})-(\ref{H13}). Write
$\psi(x,t_1)=\rho(x,t_1)^{\frac{1}{2}}\exp [\frac{i}{\hbar}S(x,t_1)]$, and define
$\tilde{\psi}(x,t):=\exp [\varphi(x,t)] \psi(x,t)$. Then, if $\varphi$ satisfies (\ref{H12}),
using the Schr\"{o}dinger equation (\ref{H7}) satisfied by $\psi$, we get
\begin{eqnarray}\nonumber&&\frac{\partial\tilde{\psi}}{\partial t}=\exp [\varphi]
\left(\frac{\partial \varphi}{\partial t}\psi+ \frac{\partial \psi}{\partial t}\right)=\\
&&-\frac{i}{\hbar}V(x)\tilde{\psi}+\frac{i\hbar}{2m}\exp\varphi\left(\Delta\psi+2\nabla\psi\cdot\nabla\varphi+
\nabla\varphi\cdot\nabla\varphi\psi+\Delta\varphi\psi\right)=
\frac{i\hbar}{2m}\Delta\tilde{\psi} -\frac{i}{\hbar}V(x)\tilde{\psi}.\nonumber \end{eqnarray}
Observing that $\tilde{\psi}(x,t_1)=\tilde{\rho}(x,t_1)^{\frac{1}{2}}\exp
[\frac{i}{\hbar}S(x,t_1)]$, we conclude that if $\tilde{\psi}$ is the solution of 
the Schr\"{o}dinger equation (\ref{H7}) on $[t_1,t_2]$ with initial condition at time $t_1$
given by $\tilde{\rho}(x,t_1)^{\frac{1}{2}}\exp [\frac{i}{\hbar}S(x,t_1)]$, then
$\varphi:=\log\frac{\tilde{\psi}}{\psi}$ solves the initial value problem
(\ref{H12})-(\ref{H13}). Thus, we get the following result. 

\begin{theorem} Suppose that $\tilde{\psi}$ is the solution of 
the Schr\"{o}dinger equation (\ref{H7}) on $[t_1,t_2]$ with initial condition at time $t_1$
given by $\tilde{\rho}(x,t_1)^{\frac{1}{2}}\exp [\frac{i}{\hbar}S(x,t_1)]$. Then 
$\varphi:=\log\frac{\tilde{\psi}}{\psi}$ solves the initial value problem
(\ref{H12})-(\ref{H13}). Suppose that $\varphi$
satisfies (\ref{H10}), and that $\tilde{\rho}_2(x)=|\tilde{\psi}(x,t_2)|^2$. Then the stochastic
process $\tilde{x}\in {\cal X}_{\tilde{\rho}_2}$ having quantum drift $\frac{\hbar}{mi}\nabla
\log\tilde{\psi}(\tilde{x}(t),t)$ solves the constrained extremization problem. \end{theorem}
Thus, by a purely probabilistic argument, we have shown that the new process after the
measurement at time $t_1$ is associated to another solution $\tilde{\psi}$ of the same
Schr\"{o}dinger equation (\ref{H7}). The association is precisely as before, namely the
quantum drift is proportional to the gradient of the logarithm of $\tilde{\psi}$. In other
words, the new process is just the Nelson process associated to the solution 
$\{\tilde{\psi}(x,t);t_1\le t\le t_2\}$. It is important to observe that the new wave function
has the same phase at time
$t_1$ as the old one before measurement. This agrees with standard quantum mechanics when it is
assumed that immediate repetition of the measurement yields the same result and does not
change the wavefunction except for an arbitrary phase factor, see e.g.
\cite{B,K}. Here, however, no further assumption is needed: {\it The invariance of the phase follows
from the variational principle}. This is a crucial point. Indeed, if we assume the invariance of the
phase after a position measurement in stochastic mechanics, then the variational principle of Section
4 suffices to produce the new Nelson process (associated to the solution  $\{\tilde{\psi}(x,t)\}$ of
the Schr\"{o}dinger equation). Also notice that the solution process possesses the same coherence
property with respect to the time interval as the solution process of the previous section.

\section{Discussion} In this paper we have shown that, in the frame of Nelson's stochastic
mechanics, the wave function reduction does not need to be {\it postulated}, but may be {\it
derived} from the standard rules of probability (Bayes' theorem) and a stochastic variational
principle of transparent significance. It seems to us that this
result lends support to the point of view of Blanchard, Golin and Serva in
\cite{BGS}, where it was shown that some apparent paradoxes of stochastic mechanics related to
repeated measurements could be removed by introducing an appropriate new process after each
measurement. The new process, indeed, is the Nelson process associated to  the new solution
$\tilde{\psi}$ of the Schr\"{o}dinger equation. A general comparison between standard quantum
mechanics and stochastic mechanics is beyond the aims of this paper, and anyway beyond the
knowledge and the understanding of the present author. We refer the reader to \cite{N2,N4}, as
well as to a series of recent papers by  Francesco Guerra
\cite{G94,G97}, for a thorough and deep analysis on the possibility of regarding Nelson's
stochastic mechanics as a complete physical theory.

Nevertheless, it seems legitimate to us to stress that stochastic mechanics, including
the elements of a theory of measurement outlined in
\cite{BGS} and here, can simply be based on the hypothesis of universal Brownian motion and on
stochastic variational principles. Thus, stochastic mechanics appears as a generalization of
classical mechanics whose foundations are completely independent from standard quantum
mechanics. Moreover, this theory is now capable of providing a transparent
probabilistic derivation of the two most mysterious features of standard quantum mechanics,
namely the uncertainty principle and the wave function collapse.

\vspace{0.4cm}
\noindent
{\bf Acknowledgement:} We wish to thank all the participants in the Fall 98 seminar 
on``Control of quantum systems", and expecially P. Marchetti, for stimulating me through very
lively discussions to (hopefully) improve my understanding on the measurement problem in
stochastic mechanics. 

\renewcommand{\baselinestretch}{1}

\end{document}